\documentclass[fl eqn,twoside]{article}
\date{}
\usepackage{amsmath}
\usepackage{amsfonts}
\usepackage{amssymb}
\usepackage{graphicx}
\def\be{\begin{equation}}
\def\ee{\end{equation}}

\begin{document}
\date{Today}
%\title{{\bf{Signature of noncommutative structure of space in resonant detectors of gravitational wave}}}
\title{{\bf{Footprint of spatial noncommutativity in resonant detectors of gravitational wave}}}

\author{
{\bf {\normalsize Sukanta Bhattacharyya}$^{a}$,\thanks{sukanta706@gmail.com}}
{\bf {\normalsize Sunandan Gangopadhyay}
$^{b}$\thanks{sunandan.gangopadhyay@gmail.com,
sunandan.gangopadhyay@bose.res.in}},
{\bf {\normalsize Anirban Saha }
$^{a}$\thanks{anirban@associates.iucaa.in}}\\
$^{a}$ {\normalsize Department of Physics, West Bengal State University, Barasat, Kolkata 700126, India}\\
$^{b}$ {\normalsize  Department of Theoretical Sciences,}\\{\normalsize S.N. Bose National Centre for Basic Sciences,}\\{\normalsize JD Block, 
Sector III, Salt Lake, Kolkata 700106, India}\\[0.1cm]
}
\date{}
\maketitle\
\begin{abstract}

\noindent
The present day gravitational wave (GW) detectors strive to detect the length variation $\delta L = h L$, which, owing to the smallness of the metric perturbation $\sim h$, is an extremely small length $\mathcal{O} \sim 10^{-18} - 10^{-21}$ meter. The recently proposed noncommutative structure of space has a characteristic length-scale $\sqrt{\theta}$ which has an estimated upper-bound in similar length-scale range. We therefore propose that GW data can be used as an effective probe of noncommutative structure of space and demonstrate how spatial noncommutativity modifies the responding frequency of the resonant mass detectors of GW and also the corresponding  probabilities of GW induced transitions that the phonon modes of the resonant mass detectors undergo. In this paper we present the complete perturbative calculation involving both time-independent and time-dependent perturbation terms in the Hamiltonian.

%\noindent It has been observed that the upper bound of the spatial noncommutative parameter is close to the variation in length measured in the gravitational wave detectors. This observation have motivated us to probe the detection of noncommutativity in spatial geometry in the gravitational wave detection scenario. We therefore investigate the two dimensional noncommutative quantum harmonic oscillator interacting with gravitational waves from various astronomical events. We treat the time independent term of noncommutative origin in the Hamiltonian of the model perturbatively and observe that this is responsible for shifting the resonance points in the quantum mechanical transitions (induced by the gravitational waves) between the ground state and excited states of the two dimensional harmonic oscillator, a feature which was absent in our previous investigation. We then incorporate the effect of the time dependent term of noncommutative origin in the Hamiltonian perturbatively. Doing this we find that the transition probabilities are with unequal intensities and that a quantum mechanical transition induced by a gravitational wave with both the $\times$ and $+$ polarizations can be an indirect evidence of the spatial noncommutative structure of  geometry, a feature which was although present in our previous study was restricted by the involvement of the $+$ polarization of the gravitational wave only.
\end{abstract}

\section{Introduction}
One of the most important achievements in the $21st$ century is the direct detection of the gravitational waves (GWs)\cite{ligo,ligo2} by the advanced LIGO detector \cite{ligoo}. 
Among the currently operating GW detectors (LIGO \cite{abramovici}, VIRGO \cite{caron}, GEO \cite{luck} and TAMA \cite{ando}) where interferometric techniques are being used, the LIGO detector has undergone new improvements and the advanced LIGO \cite{ligoo} detectors has now reached a sensitivity where one can effectively detect a length-variation of the order of $\frac{\delta L}{L} \sim 10^{-23}/\sqrt{{\rm{Hz}}} $ or better. Various resonant detectors \cite{bar_2, bar_detectors_1, bar_detectors_2, bar_detectors_3, bar_detectors_4, bar_detectors_5} are also striving to improve their sensitivity goals, specially with the new generation of spherical detectors like MiniGrail \cite{MiniGrail} and Schenberg \cite{Schenberg}. These developments are not only opening up a whole new channel for astronomical observations but also providing access to the structure of space at a length-scale resolution that has never been probed before. 

Interestingly various gedanken experiments with very high length scale resolution have predicted uncertainty in the spatial coordinates \cite{Dop, Alu} caused by a sharp localization of events in space at the quantum level. A convenient prescription to incorporate this effect into theoretical model building is to construct the quantum mechanics of the system concerned in a space that assume an underlying noncommutative (NC) geometry where the coordinate operators $\hat x_{i}$ follow  the NC Heisenberg algebra \cite{doug}
\begin{eqnarray}
\left[{\hat x}_{i}, {\hat x}_{j}\right] = i \theta_{ij}= i \theta \epsilon_{ij},~~ \left[{\hat x}_{i}, {\hat p}_{j}\right] =i \hbar \delta_{ij},~~\left[{\hat p}_{i}, {\hat p}_{j}\right] = 0 ; ~~~ i, j= 1,2
\label{1}
\end{eqnarray}
instead of the standard Heisenberg algebra of ordinary quantum mechanics
\begin{eqnarray}
\left[{\hat X}_{i}, {\hat X}_{j}\right] = 0,~~~ \left[{\hat X}_{i}, {\hat P}_{j}\right] =i \hbar \delta_{ij},~~\left[{\hat P}_{i}, {\hat P}_{j}\right] = 0~~.
\label{1h}
\end{eqnarray}
In eq.(\ref{1}), $\theta$ is the NC parameter and $ \epsilon_{ij}$ is an antisymmetric tensor. The idea of NC space gained interest in the last two decades when it was realized that the low energy effective theory of $D$-branes in the background of a Neveu-Schwarz $B$-field lives on NC space \cite{SW}. In response a wide range of theories, dubbed the NC theories, have been constructed. This includes NC quantum mechanics (NCQM) \cite{duv}-\cite{sgprl}, NC quantum field theory (NCQFT) \cite{doug} and NC gravity \cite{grav}-\cite{sgrb}. Certain possible phenomenological consequences of NC space have also been predicted in \cite{jabbari1}-\cite{sun}. Naturally a part of the endeavor has also been spent in finding the order of magnitude of the NC parameter and exploring its connection with observations \cite{mpr}-\cite{stern}. The upperbound on the coordinate commutator $|\theta|$ found in \cite{carol} is $\approx \left(10 {\rm TeV}\right)^{-2}$ which corresponds to a length scale $\sqrt{4 \times 10^{-40} {\rm m}^{2}}$ for $\hbar$$=$$c$$=$$1$\footnote{In a more general NC spacetime structure \cite{SW} given by $\left[x^{\mu}, x^{\nu}\right] = i \theta^{\mu \nu}$, such upperbounds on time-space NC parameter is $\theta^{0i}$ $\approx 9.51\times 10^{-18} {\rm m}^{2}$ \cite{ani}.}. However, there are studies in NCQM that suggests that the NC parameter associated with different particles may not be the same \cite{pmho, vassilavich} and this bound could be as high as $\theta \approx \left(4 {\rm GeV}\right)^{-2} - \left(30 {\rm MeV}\right)^{-2}$ \cite{stern}.
Looking at these numbers %$\Delta L \sim 10^{-19} {\rm m}$, 
it is not far fetched to anticipate that a good possibility of detecting the NC structure of space would be in the GW detection experiments as it may as well pick up the NC signature. 

\noindent Though mainstream endeavor of GW detection has shifted its focus to the more efficient interferometric detection of GW, chronologically the idea of observing GWs was pioneered by J. Weber who designed the resonant bar detectors \cite{Weber_1}-\cite{Weber_3} in 1960's, where GW causes the phonon modes of the bar to oscillate resonantly. This resonance turns the energy carried by the GW to the mechanical energy of the bar which (by using transducer attached to the bar) is transformed to electrical energy that is in turn detected. The study of these resonant bar detectors is still interesting and fundamental because it focuses on how GW interacts with elastic matter causing vibrations with amplitudes many order smaller than the size of a nucleus. In a present day bar detector \cite{GW-detection_status} it is possible to detect variations $\Delta L$ of the bar-length $L \sim 1 {\rm m}$, with $\frac{\Delta L}{L} \sim 10^{-19}$. The tiny vibrations called phonons \cite{Magg} are nothing but quantum mechanical forced harmonic oscillators (HO). Thus the response of a resonant detector to GW can be quantum mechanically described as GW-HO interaction.

Now if the spatial structure at the quantum level is inherently noncommutative in nature, a quantum mechanical theory of the GW-HO interaction formulated in NC space, that is NCQM would be necessary to predict the possible NC effects in the GW detector read-outs. For this purpose we had earlier studied various aspects of the GW-HO interaction in NCQM framework in \cite{ncgw1}-\cite{prd}. In these studies we worked out the formal solution to the system which show that the spatial noncommutativity introduces a characteristic shift in the frequency value where the HO will resonate with the GW. In \cite{cqg}, we computed the transition probabilities between the ground state and the excited states of this system where only the time-dependent perturbation terms were taken into consideration. That work \cite{cqg}, though demonstrated significant effect of noncommatativity in the transition probabilities, could not capture the aforesaid NC shift in the resonance point since that is connected with the removal of degeneracy of excited states and perturbative correction of energy levels of the HO by the time-independent part of the perturbation. In this paper, we therefore incorporate the complete perturbative computation. Interestingly, we observe that this inclusion not only captures the characteristic shift in the resonant frequency but also significantly modifies the transition probabilities and helps to draw newer conclusions regarding the possibility of detecting NC effect in GW detection data. 

The organization of the paper is as follows. In section 2 we briefly outline how the HO-GW interaction can be modeled in a noncommutative space and obtain the relevant Hamiltonian. The complete perturbative calculation to obtain the working formula for transition probabilities among the shifted energy levels for a generic GW wave-form is presented in section 3. In section 4 we use various GW wave-forms to calculate the corresponding transitions probabilities and discuss the possibilities of detecting NC signature as a consequence. We conclude in section 5.
%%%%%%%%%%%%%%%%%%%%%%
\section{The NC HO-GW interaction model}
%%%%%%%%%%%%%%%%%%%%%%

In a bar detector the incoming GW interacts with the phonon mode excitations which can be described as quantum mechanical forced harmonic oscillators. This allows us to model the system as GW-HO interaction and since the effect of GW is constrained to the plane perpendicular to the propagation vector of the GW (taken in $z$-direction in this work), we consider a $2-$dimensional HO in the $x-y$ plane. To start with, we first need to have the classical Lagrangian describing the  same. This reads upto a total derivative as
\begin{equation}
{\cal L} = \frac{1}{2} m\left(\dot {x_{j}}\right)^2 - m{\Gamma^j}_{0k}
\dot {x}_{j} {x}^{k}  - \frac{1}{2} m \varpi^{2} \left(x_{j}\right)^2
\label{e8}
\end{equation}
where ${R^j}_{0,k0} = - \frac{d \Gamma^j_{0k}}{d t}  = -\ddot{h}_{jk}/2  $,  ${R^j}_{0,k0}$ denotes the components of the curvature tensor in terms of the metric perturbation $h_{\mu \nu} $ as
\begin{eqnarray}
g_{\mu\nu} = \eta_{\mu\nu} + h_{\mu\nu}~; \, |h_{\mu\nu}|<<1
\label{metric_perturbation}
\end{eqnarray}
on the flat Minkowski background $\eta_{\mu\nu}$. The Lagrangian (\ref{e8}) leads to the geodesic deviation equation for a $2-$dimensional HO of mass $m$ and intrinsic frequency $\varpi$ in a proper detector frame that reads 
\begin{equation}
m \ddot{{x}} ^{j}= - m{R^j}_{0,k0} {x}^{k} - m \varpi^{2} x^{j}
\label{e5}
\end{equation}
where dot denotes derivative with respect to the coordinate time of the proper detector frame\footnote{It is the same as it's proper time to first order in the metric perturbation.}, 
and ${x}^{j}$ is the proper distance of the pendulum from the origin. The validity of eq.(\ref{e5}) remains only in the small velocity and long wavelength limit\footnote{Small velocity implies that the spatial velocities involved are non-relativistic and long wavelength implies that $|{x}^{j}|$ is much smaller than the reduced wavelength $\frac{\lambda}{2\pi}$ of the GW.}. These conditions are obeyed by the resonant bar detectors and earth bound interferometric detectors. It is to be noted that we have already imposed the transverse-traceless (TT) gauge to remove the unphysical degrees of freedom and hence the only relevant components of the curvature tensor ${R^j}_{0,k0} = -\ddot{h}_{jk}/2$ arises in eq.(\ref{e5}). The TT gauge choice gives only two physical degrees of freedom, namely, the $\times$ and $+$ polarizations of the GW.
%This also ensures that in a plane-wave expansion of GW, $h_{jk} = \int  (A_{jk} e^{ikx} + A^{*}_{jk} e^{- ikx}) d^{3}k/\left(2 \pi\right)^{3},$ the spatial part $e^{i \vec{k}.\vec{x}} \approx 1$ all over the detector site. Thus our only concern is the time-dependent part of the GW. If the polarization information contained in $A_{jk}$ is expressed in terms of the Pauli spin matrices, $h_{jk}$ takes the most convenient form 
The most convenient form of $h_{jk}$ explicitly showing these polarizations can be written in terms of the Pauli spin matrices $\sigma^1$ and $\sigma^3$ and reads
\begin{equation}
h_{jk} \left(t\right) = 2f \left(\varepsilon_{\times}\sigma^1_{jk} + \varepsilon_{+}\sigma^3_{jk}\right)
\label{e13}
\end{equation}
where $2f$ is the amplitude of the GW and $\left( \varepsilon_{\times}, \varepsilon_{+} \right)$ are the two possible polarization states of the GW satisfying the condition $\varepsilon_{\times}^2+\varepsilon_{+}^2 = 1$ for all $t$. In case of linearly polarized GW, the frequency $\Omega$ is contained in the time dependent amplitude $2f(t)$ whereas the time dependent polarization states $\left( \varepsilon_{\times} \left(t \right), \varepsilon_{+} \left( t \right) \right)$ contains the frequency $\Omega$ for the circularly polarized GW.

\noindent The canonical momentum corresponding to ${x}_{j}$ for the HO is ${p}_{j} = m\dot {x}_{j} - m \Gamma^j_{0k} {x}^{k}$ and it gives the Hamiltonian by a Legendre transformation 
\begin{equation}
{H} = \frac{1}{2m}\left({p}_{j} + m \Gamma^j_{0k} {x}^{k}\right)^2 + \frac{1}{2} m \varpi^{2} x_{j}^2 ~.
\label{e9}
\end{equation}
With this set up in place, we now move on to investigate the NC quantum mechanical description of the system. This is done by simply elevating the phase-space variables $\left( x^{j}, p_{j} \right)$ to operators $\left( {\hat x}^{j}, {\hat p}_{j} \right)$ and imposing the NC Heisenberg algebra. This algebra can be mapped to the standard $\left( \theta = 0 \right)$ Heisenberg algebra spanned by the operators $X_{i}$ and $P_{i}$ of ordinary QM through \cite{cst, stern}
\begin{eqnarray}
{\hat x}_{i} = X_{i} - \frac{1}{2 \hbar}
\theta \epsilon_{ij} P_{j}~, \quad {\hat p}_{i} = P_{i}~.
\label{e9b}
\end{eqnarray}
With this map, the Hamiltonian in eq.(\ref{e9}) can be written in terms of the commutative variables as\footnote{The traceless property of the GW is required here.}
\begin{eqnarray}
{\hat H} = \left(\frac{ P_{j}{}^{2}}{2m} + \frac{1}{2} m \varpi^{2} X_{j}{}^{2} \right)+ \Gamma^j_{0k} X_{j} P_{k} -\frac{m \varpi^{2}}{2 \hbar} \theta \epsilon_{jm} X^{j} P_{m} 
%\nonumber \\ &&
-\frac{\theta }{2 \hbar} \epsilon_{jm} P_{m} P_{k}  \Gamma^j_{0k} + \mathcal O(\Gamma^2) ~.
\label{e12}
\end{eqnarray}
In eq.(\ref{e12}) the first-bracketed terms represent the unperturbed HO Hamiltonian $\hat{H}_{0}$ while all the remaining terms are small compared to $\hat{H}_{0}$ and can be treated as perturbations. Third term is a pure HO-GW interaction term and the remaining two terms bear the effect of noncommutativity. 

\noindent We now define raising and lowering operators in terms of the oscillator frequency $\varpi$
\begin{eqnarray}
X_j = \sqrt{ \frac{\hbar}{2m \varpi}}\left(a_j+a_j^\dagger\right); \quad
%\label{e15a} \\
P_j = \sqrt{ \frac{\hbar m\varpi}{2i}}
\left(a_j-a_j^\dagger\right)\ .
\label{e15}
\end{eqnarray}
The Hamiltonian  in terms of raising and lowering operator can be recast as
\begin{eqnarray}
{\hat H} &=&\hbar \varpi (a_j^\dagger a_j+1) - \frac{i\hbar}{4} \dot h_{jk}(t) \left(a_j a_k - a_j^\dagger a_k^\dagger\right)
+ \frac{m \varpi \theta}{8} \epsilon_{jm} {\dot h}_{jk}(t)  \left(a_{m}a_{k}  - a_{m}a_{k}^\dagger + C.C. \right) \nonumber\\ &&
-\frac{i}{2} m \varpi^2 \theta \epsilon_{jk} a_j^\dagger a_k  \equiv \hat{H}_{0} + \hat{H}_{{\rm {int}}} (t)
\label{e16}
\end{eqnarray}
where C.C. means complex conjugate. In eq.(\ref{e16}) the first term represents $\hat{H}_{0}$, the standard HO and all the perturbation terms are collected in $\hat{H}_{{\rm {int} }}(t)$. In the next section we proceed with the perturbative calculations.

%%%%%%%%%%%%%%%%%%%%%%%%
\section{Perturbed energy levels and transitions}
%%%%%%%%%%%%%%%%%%%%%%%%
Our aim in this section is to calculate the perturbed states, their corresponding energy levels and transition probabilities among them. For this we first break the interaction Hamiltonian $\hat{H}_{{\rm {int} }}(t)$ as follows
\begin{eqnarray}
\hat{H}_{\rm {int}} (t) = \hat{H_1} + \hat{H_2}(t)~.
\label{e17}
\end{eqnarray}
%%%%%%%%%%%%%%%%%%%%%%
The time independent perturbation shifting the energy levels is
\begin{eqnarray}
\hat{H}_1 = - i \Lambda_{\theta} \hbar \epsilon_{jk} a_j^\dagger a_k ~.
\label{h1} 
\end{eqnarray}
Notice that this is a purely NC term with no effect of GW in it and it also introduces a characteristic frequency of noncommutativity
\begin{eqnarray}
\Lambda_{\theta} =  \frac{m \varpi^2 \theta}{2 \hbar}
\label{Lambda_theta}
\end{eqnarray}
which depends on the mass $m$ and intrinsic frequency $\varpi$ of the HO. The time dependent perturbations which induce the transitions are given by
\begin{eqnarray}
\hat{H}_2(t) = - \frac{i\hbar}{4} \dot h_{jk}(t) \left(a_j a_k - a_j^\dagger a_k^\dagger\right) + \frac{\Lambda}{4} \hbar \epsilon_{jm} {\dot h}_{jk}(t)  \left(a_{m}a_{k}  - a_{m}a_{k}^\dagger + C.C. \right) 
\label{e18}
\end{eqnarray}
where 
\begin{eqnarray}
\Lambda= \frac{m \varpi \theta}{2 \hbar} 
\label{Lambda}
\end{eqnarray}
is a dimensionless parameter that is again a characteristic of noncommutativity and gives a measure of how strongly the NC structure of space affects the transitions\footnote{This will be explicitly demonstrated in the results.}. Also note that the characteristic NC frequency can also be written as $\Lambda_{\theta} =\varpi \Lambda$.

\noindent We first calculate the perturbed energy states incorporating the effect of $\hat{H_1}$ using time independent perturbation theory. It turns out that the spatial NC structure removes the degeneracy of the second excited state of the $2$-dimensional HO as expected from the time independent degenerate perturbation theory to yield the following perturbed eigenstates  
\begin{eqnarray}
\psi_2^{(0)} &=& (| 2,0\rangle + | 0,2\rangle)\nonumber\\
\psi_2^{(1)} &=& (| 2,0\rangle - | 0,2\rangle + i \sqrt{2} | 1,1\rangle ) \nonumber\\
\psi_2^{(2)} &=& (| 2,0\rangle - | 0,2\rangle - i \sqrt{2} | 1,1\rangle )
\label{f1}
\end{eqnarray}
with the the corresponding energy eigenvalues  
\begin{eqnarray}
E_2^{(0)} &=& 3 \hbar \varpi \nonumber\\
E_2^{(1)} &=& 3 \hbar \varpi (1 + \frac{2}{3} \Lambda) \nonumber\\
E_2^{(2)} &=& 3 \hbar \varpi (1 - \frac{2}{3} \Lambda)~.
\label{ev}
\end{eqnarray}
As has been mentioned earlier we see that the relative shift of the energy level depends on the size of the dimensionless NC parameter $\Lambda$.

We now proceed to include the effect of the time dependent perturbation $\hat{H_2}\left(t\right)$ to compute the transition probabilities between the ground state and the perturbed non-degenerate excited states of the $2$-dimensional harmonic oscillator using time dependent perturbation theory.
Now to the lowest order of approximation in time dependent perturbation theory, the probability amplitude of transition from an initial state $|i\rangle$ to a final state $|f \rangle$, ($i\neq f$), due to a perturbation $\hat{V}(t)$ is given by \cite{kurt}
\begin{eqnarray}
C_{i \rightarrow f}(t\rightarrow \infty)  =  -\frac{i}{\hbar} \int_{-\infty}^{t\rightarrow +\infty} dt' F \left( t' \right) e^{\frac{i}{\hbar}(E_f -E_i)t'} \langle \Phi_f | \hat{Q}|\Phi_i \rangle
\label{probamp}
\end{eqnarray}
where $\hat{V}(t)=F(t)\hat{Q}$. Using the above result, we observe that the probability of transition survives only between the ground state $|0,0\rangle$ and the perturbed states given by eq.$(\ref{f1})$:
\begin{eqnarray}
C_{0\rightarrow 2^{(0)}} = - \frac{i}{\hbar} \int_{-\infty}^{+\infty} dt && \left[ F_{jk} \left( t \right) e^{\frac{i}{\hbar} \left( E_{2}^0 - E_{0}\right)t} \left (\langle 2,0|\hat Q_{jk}|0,0\rangle \right. \right. \left. \left. + \langle 0,2|\hat Q_{jk}|0,0\rangle\right)\right] \nonumber\\
C_{0\rightarrow 2^{(1)}} = - \frac{i}{\hbar} \int_{-\infty}^{+\infty} dt && \left[ F_{jk} \left( t \right) e^{\frac{i}{\hbar} \left( E_{2}^1 - E_{0}\right)t} \left (\langle 2,0|\hat Q_{jk}|0,0\rangle \right. \right. \left. \left.+i\sqrt{2} \langle 1,1|\hat Q_{jk}|0,0\rangle - \langle 0,2|\hat Q_{jk}|0,0\rangle\right)\right] \nonumber\\
C_{0\rightarrow 2^{(2)}} = - \frac{i}{\hbar} \int_{-\infty}^{+\infty} dt && \left[ F_{jk} \left( t \right) e^{\frac{i}{\hbar} \left( E_{2}^2 - E_{0}\right)t} \left (\langle 2,0|\hat Q_{jk}|0,0\rangle \right. \right. \left. \left. - i\sqrt{2} \langle 1,1|\hat Q_{jk}|0,0\rangle - \langle 0,2|\hat Q_{jk}|0,0\rangle\right)\right]
\label{trans_amp_02}
\end{eqnarray}
where $F_{jk} \left( t \right) = \dot h_{jk}(t)$ contains the explicit time dependence of ${\hat {H_2}}(t)$ and
\begin{eqnarray}
\hat Q_{jk} &=& - \frac{i\hbar}{4} \left(a_j a_k - a_j^\dagger a_k^\dagger\right) + \frac{\Lambda}{4}  \hbar \epsilon_{jm}  \left( a_{m}a_{k}  - a_{m}a_{k}^\dagger + C.C. \right).
\label{Q}
\end{eqnarray}
Expanding out ${\hat Q_{jk}}$ for $j,k = 1,2$, we obtain the transition amplitude between the ground state $|0,0\rangle$ and the second excited state to be
\begin{eqnarray}
C_{0\rightarrow 2^{(0)}} &=& 0 \nonumber\\
C_{0\rightarrow 2^{(1)}} &=& - \frac{i}{\hbar} \int_{-\infty}^{+\infty} dt \, e^{2i \varpi(1 + \Lambda)t} \hbar \left[ i  A(\Lambda) \dot h_{11}(t) - B(\Lambda) \dot h_{12}(t)\right].\nonumber\\
\rm  \nonumber\\
C_{0\rightarrow 2^{(2)}} &=& - \frac{i}{\hbar} \int_{-\infty}^{+\infty} dt \, e^{2 i \varpi(1 - \Lambda)t}  \hbar \left[ i  C(\Lambda) \dot h_{11}(t) - D(\Lambda) \dot h_{12}(t)\right]
\label{trans_amp_02a}
\end{eqnarray}
where
\begin{eqnarray}
A(\Lambda)=\frac{1}{\sqrt{2}} \left(1+ \Lambda \right), \quad B(\Lambda)=\frac{1}{\sqrt{2}} \left(\sqrt{\frac{3}{2}} \Lambda  + 1 \right)~, \quad
C(\Lambda)=\frac{1}{\sqrt{2}} \left(1 - \Lambda \right)~, \quad
D(\Lambda)=\frac{1}{\sqrt2} \left(\sqrt{\frac{3}{2}} \Lambda - 1 \right) .
\label{dim_less_NC}
\end{eqnarray}
\noindent Eq.(\ref{trans_amp_02a}) is the main working formula in this paper. In the next section
we use the general formula (\ref{trans_amp_02a}) and compute the corresponding transition probabilities from the relation
\begin{eqnarray}
P_{0\rightarrow 2} =  |C_{0\rightarrow 2}|^{2}~~.
\label{trans_prob}
\end{eqnarray}

%%%%%%%%%%%%%%%%%%%%%%%%%%%%%
\section{Transition probabilities for different types of gravitational wave}
%%%%%%%%%%%%%%%%%%%%%%%%%%%%%
In this section we take various templates of gravitational wave-forms that are likely to be generated in runaway astronomical events and calculate the transition probabilities due to these gravitational wave-forms.
%%%%%%%%%%%%%%%%%%%%%%
\subsection{Periodic linearly polarized GW}
%%%%%%%%%%%%%%%%%%%%%%
\noindent First we consider the simple scenario of periodic GW with linear polarization. This can be written as
\begin{equation}
h_{jk} \left(t\right) = 2f_{0} \cos{\Omega t} \left(\varepsilon_{\times}\sigma^1_{jk} + \varepsilon_{+}\sigma^3_{jk}\right)
\label{lin_pol}
\end{equation}
where the amplitude varies sinusoidally with a single frequency $\Omega$. In this case, we get the transition probabilities to be
\begin{eqnarray}
P_{0\rightarrow 2^{(1)}} &=&  \left( \pi f_{0} \Omega \right)^{2}  \left[A(\Lambda)^2 \varepsilon_{+}{}^{2}  + B(\Lambda)^2 \varepsilon_{\times}{}^{2} \right] \times
\left[\delta \left(2 \varpi_{+} + \Omega \right) - \delta \left(2  \varpi_{+} - \Omega \right) \right]^{2} \nonumber\\
\rm \nonumber\\
P_{0\rightarrow 2^{(2)}} &=&  \left( \pi f_{0} \Omega \right)^{2}  \left[C(\Lambda)^2 \varepsilon_{+}{}^{2}  + D(\Lambda)^2 \varepsilon_{\times}{}^{2}   \right] \times
\left[\delta \left(2 \varpi_{-} + \Omega \right) - \delta \left(2 \varpi_{-} - \Omega \right) \right]^{2}
\label{trans_prob_lin_pol}
\end{eqnarray}
\noindent where $\varpi_{+}$ and $\varpi_{-}$ are given by
\begin{eqnarray}
\varpi_{+} = \varpi + \Lambda_{\theta}= \varpi (1 + \Lambda) ,\quad \varpi_{-} =   \varpi - \Lambda_{\theta} = \varpi (1- \Lambda).
\label{varpi}
\end{eqnarray}
\noindent Owing to the restriction of the physical range of frequency $\left(0< \varpi < \infty \right) $, the delta functions $\delta \left(2 \varpi_{+} + \Omega \right)$ and $\delta \left(2 \varpi_{-} + \Omega \right)$ in the transition probabilities in eq.(\ref{trans_prob_lin_pol}) do not contribute. Hence, we have
\begin{eqnarray}
P_{0\rightarrow 2^{(1)}} &=&  \left( \pi f_{0} \Omega \right)^{2}  \left[A(\Lambda)^2 \varepsilon_{+}{}^{2}  + B(\Lambda)^2 \varepsilon_{\times}{}^{2} \right] \times
\left[\delta \left(2 \varpi_{+} - \Omega \right) \delta (0) \right] \nonumber\\
\rm  \nonumber\\
P_{0\rightarrow 2^{(2)}} &=&  \left( \pi f_{0} \Omega \right)^{2}  \left[C(\Lambda)^2 \varepsilon_{+}{}^{2}  + D(\Lambda)^2 \varepsilon_{\times}{}^{2}   \right] \times
\left[\delta \left(2 \varpi_{-} - \Omega \right) \delta(0) \right]~.
\label{prob_res}
\end{eqnarray}
Now noting that in any real experimental situation, the GW signal is received for a finite time interval $T$, we regularize the Dirac delta function for a finite observation time $-\frac{T}{2}<t<\frac{T}{2}$ as
\begin{eqnarray}
\delta(\varpi)=\left[ \int_{-\frac{T}{2}}^{\frac{T}{2}} dt \, e^{i \varpi t}\right] = T~.
\label{time period}
\end{eqnarray}
\noindent The transition rates then take the form 
\begin{eqnarray}
\lim\limits_{T \rightarrow \infty} \frac{1}{T}P_{0\rightarrow 2^{(1)}} &= & \left( \pi f_{0} \Omega \right)^{2}  \left[A(\Lambda)^2 \varepsilon_{+}{}^{2}  + B(\Lambda)^2 \varepsilon_{\times}{}^{2} \right] \times \delta \left(2 \varpi_{+} - \Omega \right) 
\label{tm} \\
\lim\limits_{T \rightarrow \infty} \frac{1}{T}P_{0\rightarrow 2^{(2)}} &=&  \left( \pi f_{0} \Omega \right)^{2}  \left[C(\Lambda)^2 \varepsilon_{+}{}^{2}  + D(\Lambda)^2 \varepsilon_{\times}{}^{2}   \right] \times \delta \left(2 \varpi_{-} - \Omega \right)~.
\label{trans_rate}
\end{eqnarray}
%Let us point out the salient features of this result
From the above results we can make following observations:
\begin{enumerate}
\item
The delta functions in eq.(s) (\ref{tm}) and (\ref{trans_rate}) ensure that the transition rates will be peaked around the frequencies $\Omega= 2 \varpi_{+}$ and $\Omega= 2 \varpi_{-}$. Observationally speaking the resonance occurs when the frequencies of the GW ($\Omega $) matches with that of the resonant bar detector. Thus looking at the expressions of $ \varpi_{+}$ and $ \varpi_{-}$ in eq.(\ref{varpi}) we find that {\it we should get two resonant points if the space has an underlying NC geometry} instead of the single resonant point at $\Omega= 2 \varpi$ which is expected otherwise.
\item 
Using $A, B, C, D$ from eq.(\ref{dim_less_NC}) in the expressions for the transition probabilities in eq.(s) (\ref{tm}) and (\ref{trans_rate}) it is easy to see that that the transition probability $P_{0\rightarrow 2^{(1)}}$ is larger than $P_{0\rightarrow 2^{(2)}}$. Hence the two transitions lines at frequencies $\Omega= 2\varpi_{+}$ and $\Omega= 2\varpi_{-}$ are {\it not of equal strength.} This is another characteristic feature which, if present in the bar detector data, will signify the presence of spatial noncommutativity.
\item
From the expressions of $A, B, C, D$ in eq.(\ref{dim_less_NC}) it is clear that terms linear and quadratic in the dimensionless NC parameter $\Lambda$ will appear in the transition probabilities (\ref{tm}) and (\ref{trans_rate}). This is a welcome result since experimentally linear dependence is easier to observe for small values of $\Lambda$.
The value of $\Lambda$ depends crucially on how the quantum mechanical HO is realized. Shortly we shall demonstrate that in the context of resonant bar detectors this value can be in principle detectable.
\item 
Further, in \cite{cqg} it was observed that only for the $+$ polarization of the GW, the transition probability contained the effect of the dimensional NC parameter $\Lambda$. But here we find that both the $+$ and the $\times$ polarizations includes the effects of the NC structure of space. This is a very interesting feature that arises in our analysis and was absent in our earlier investigation \cite{cqg}. 
\end{enumerate}
With these observations in place we now move on to compute the transition probabilities for circularly polarized GW and investigate whether the same holds.
\subsection{Periodic circularly polarized GW}
%%%%%%%%%%%%%%%%%%%%%%%
\noindent  The simplest form of a periodic GW signal with circular polarization  can be conveniently expressed as
\begin{equation}
h_{jk} \left( t \right) = 2f_{0} \left[\varepsilon_{\times} \left( t \right) \sigma^1_{jk} + \varepsilon_{+}\left( t \right) \sigma^3_{jk}\right] 
\label{cir_pol}
\end{equation}
with $\varepsilon_{+} \left( t \right)  = \cos \Omega t $ and $\varepsilon_{\times} \left( t \right)  = \sin \Omega t $ and $\Omega$ is the frequency of GW. The transition probabilities in this case are given by 
\begin{eqnarray}
P_{0\rightarrow 2^{(1)}} &=& \left ( \frac{ f_{0} \Omega}{\hbar} \right)^{2} \times \left [A(\Lambda)^2 \big(\delta(2\varpi_{+} + \Omega) + \delta(2 \varpi_{+} - \Omega)\big )^{2} +  B(\Lambda)^2 \big(\delta (2\varpi_{+} + \Omega) - \delta(2 \varpi_{+} - \Omega)\big)^{2} \right ]
\label{tp} \\
P_{0\rightarrow 2^{(2)}} &=& \left ( \frac{ f_{0} \Omega}{\hbar} \right)^{2} \times \left [C(\Lambda)^2 \big(\delta(2 \varpi_{-} + \Omega) + \delta(2 \varpi_{-} - \Omega)\big)^{2} +  D(\Lambda)^2 \big(\delta(2 \varpi_{-} + \Omega) - \delta(2 \varpi_{-} - \Omega)\big)^{2} \right ].
\label{trans_prob_cir_pol}
\end{eqnarray}
\noindent Once again imposing the physical restriction of the natural frequency of the detector ($ 0<\varpi< \infty$) and regularizing the Dirac delta function for a finite observation time $-\frac{T}{2}<t<\frac{T}{2}$ (as has been done in the last section) the transition rates for a real experimental situation become
\begin{eqnarray}
\lim\limits_{T \rightarrow \infty} \frac{1}{T} P_{0\rightarrow 2^{(1)}} &=& \left ( \frac{ f_{0} \Omega}{\hbar} \right)^{2} \times \left [A(\Lambda)^2 +  B(\Lambda)^2  \right ] \delta \big(2 \varpi_{+} - \Omega \big)
\label{tp1}
\\
\lim\limits_{T \rightarrow \infty} \frac{1}{T} P_{0\rightarrow 2^{(2)}} &=& \left ( \frac{ f_{0} \Omega}{\hbar} \right)^{2} \times \left [C(\Lambda)^2 +  D(\Lambda)^2 \right ] \delta \big(2 \varpi_{-} - \Omega \big)~.
\label{trans_prob_cir_pol1}
\end{eqnarray}
Note that observations 1, 2 and 3 that we made in the last section about the transition rates for linearly polarized GW holds for circularly polarized GW signals as well. Having checked that our observations hold good for periodic GW signals let us now move on to consider the case of aperiodic GW signals in the next section.

%From the above results one can easily identify the same characteristic features which can act as probe of  spatial noncommutativity in GW data for circularly polarized GW.

%We once again observe that the transitions occur when the NC modified natural frequency of the harmonic oscillator resonates with the frequency of the GW source. Further, there are two different resonant points with unequal transition probability intensities. It is reassuring to note that there is only one transition point instead of two (with unequal heights) when $\Lambda=0$. {\it Thus we can also use  circularly polarized GW from a binary system as a deterministic probe for spatial noncommutativity.} This result differs from that in \cite{cqg} due to the contribution of the time independent perturbative term of NC origin in the Hamiltonian (\ref{e16}).

%%%%%%%%%%%%%%%%%%%%%%%%%%
\subsection{Aperiodic linearly polarized GW: Burst}
%%%%%%%%%%%%%%%%%%%%%%%%%%
\noindent Common examples of GW with aperiodic signals are GW bursts. Such signals are expected from  inspiraling neutron stars or black hole binaries.  At their last stable orbit or during their merger and final ringdown, they emit GW signal with a huge amount of energy within a very short duration $10^{-3} {\rm sec} < \tau_{{\rm g}} < 1 {\rm sec}$. Bursts originating from such violent and explosive astrophysical phenomena can only be approximately modeled and we take a simple choice as the following
\begin{eqnarray}
h_{jk} \left(t\right) = 2f_{0} g \left( t \right) \left(\varepsilon_{\times}\sigma^1_{jk} + \varepsilon_{+}\sigma^3_{jk}\right)
\label{lin_pol_burst}
\end{eqnarray}
containing both components of linear polarization. The smooth function $g \left( t \right)$  needs to go to zero rather fast for $|t| > \tau_{{\rm g}} $. Let us take a Gaussian form for the function $g(t)$
\begin{equation}
g \left(t\right) = e^{- t^{2}/ \tau_{g}^{2}}
\label{burst_waveform_Gaussian}
\end{equation}
with $\tau_g \sim \frac{1}{f_{max}}$, where $f_{max}$ is the maximum value of a broad range continuum spectrum of frequency. The burst contains such a wide range of frequency due to its small temporal duration \cite{cqg}.  
Note that at $t=0$, $g \left( t \right)$ goes to unity.
Now the Fourier decomposed modes of the GW burst can be written as 
\begin{eqnarray}
h_{jk} \left(t\right) = \frac{f_{0}}{\pi} \left(\varepsilon_{\times}\sigma^1_{jk} + \varepsilon_{+}\sigma^3_{jk}\right)  \int_{-\infty}^{+\infty} \tilde{g} \left( \Omega \right) e^{- i \Omega t}  d \Omega 
\label{lin_pol_burst_Gaussian}
\end{eqnarray}
where $\tilde{g} \left( \Omega \right) = \sqrt{\pi} \tau_{g} e^{- \left( \frac{\Omega \tau_{g}}{ 2} \right)^{2}}$ is the amplitude of the Fourier mode at frequency $\Omega$.

\noindent Using eq.(\ref{lin_pol_burst_Gaussian}) in the general formula for the transition amplitude (\ref{trans_amp_02a}) from the ground state to the second excited state, we obtain the transition amplitudes induced by a GW burst to be
\begin{eqnarray}
C_{0\rightarrow 2^{(1)}}&=& \frac{f_0}{\pi} \int_{-\infty}^{+\infty} d\Omega \left[ \Omega \tilde{g} \left( \Omega \right) \big( i A(\Lambda)  \varepsilon_{+}- B(\Lambda)  \varepsilon_{\times} \big) \int_{-\infty}^{+\infty} dt~~ e^{i(2\varpi_{+}-\Omega)t}\right] \nonumber\\
&=&2 f_0 \int_{-\infty}^{+\infty} d\Omega \left[ \Omega \tilde{g} \left( \Omega \right) \big( i A(\Lambda)  \varepsilon_{+}- B(\Lambda)  \varepsilon_{\times} \big)\delta(2 \varpi_{+} - \Omega)\right]\nonumber\\
&=& 4 f_0 \varpi_{+} \tilde{g} \left( 2 \varpi_{+} \right) \big( i A(\Lambda)  \varepsilon_{+}- B(\Lambda)  \varepsilon_{\times} \big)
\label{gwbta1}
\end{eqnarray}
\begin{eqnarray}
C_{0\rightarrow 2^{(2)}}&=& \frac{f_0}{\pi} \int_{-\infty}^{+\infty} d\Omega \left[ \Omega \tilde{g} \left( \Omega \right) \big( - i C(\Lambda)  \varepsilon_{+} + D(\Lambda)  \varepsilon_{\times} \big) \int_{-\infty}^{+\infty} dt~~ e^{i(2\varpi_{-}-\Omega)t}\right]\nonumber\\
&=& 2 f_0 \int_{-\infty}^{+\infty} d\Omega \left[ \Omega \tilde{g} \left( \Omega \right) \big( - i C(\Lambda)  \varepsilon_{+} + D(\Lambda)  \varepsilon_{\times}\big) \delta(2 \varpi_{-} - \Omega)\big) \right]\nonumber\\
&=& 4 f_0 \varpi_{-} \tilde{g}( 2 \varpi_{-}) \big( -i C(\Lambda)  \varepsilon_{+} + D(\Lambda)  \varepsilon_{\times} \big).
\label{gwbta2}
\end{eqnarray}
Now using the expression of $\tilde{g}(\Omega)$, the transition amplitudes in eq.(s)(\ref{gwbta1}, \ref{gwbta2}) simplifies to
\begin{eqnarray}
C_{0\rightarrow 2^{(1)}}&=&\left( 4 \sqrt{\pi}  f_{0} \tau_{g} \varpi_{+} \right) e^{- \tau_{g}^2 \varpi_{+}^2} \left[ i A(\Lambda) \varepsilon_{+}  - B(\Lambda) \varepsilon_{\times}  \right]
\label{gwb1}
\end{eqnarray}
\begin{eqnarray}
C_{0\rightarrow 2^{(2)}}&=  &\left( 4   \sqrt{\pi}  f_{0} \tau_{g} \varpi_{-}\right) e^{-\tau_{g}^2 \varpi_{-}^2 } \left[ C(\Lambda) \varepsilon_{+}  + i D(\Lambda) \varepsilon_{\times}  \right].
\label{gwb2}
\end{eqnarray}
We are now ready to write down the transition probabilities between the states. 
Using eq.(\ref{trans_prob}), the transition probabilities take the form 
\begin{eqnarray}
P_{0\rightarrow 2^{(1)}} & = &\left[ 4 \sqrt{\pi}  f_{0} \tau_{g} \varpi_{+} \right]^{2} e^{- 2 \tau_{g}^2 \varpi_{+}^2} \left( A(\Lambda)^2 \varepsilon_{+}{}^{2}  + B(\Lambda)^{2} \varepsilon_{\times}{}^{2}  \right)
\label{trans_prob_Gaussian_burst}\\
P_{0\rightarrow 2^{(2)}} & = &\left[ 4  \sqrt{\pi}  f_{0} \tau_{g} \varpi_{-}\right]^{2} e^{-2 \tau_{g}^2 \varpi_{-}^2 } \left( C(\Lambda)^2 \varepsilon_{+}{}^{2}  + D(\Lambda)^{2} \varepsilon_{\times}{}^{2}  \right).
\label{trans_prob_Gaussian_burst1}
\end{eqnarray}
From eq.(s) (\ref{trans_prob_Gaussian_burst}, \ref{trans_prob_Gaussian_burst1}), we once again observe that contribution in the transition probability induced by both the $+$ and the $\times$ polarized part of the GW signal are affected by spatial noncommutativity. Further the intensity of transition lines are different for the two degeneracy-lifted second excited states. Also we notice that terms both  linear and quadratic in the NC parameter $\Lambda$ comes in the transition probability. The linear dependence may be important in case the value of $\Lambda$ is small in certain realization of quantum mechanical HO, as we have discussed earlier.

%{\it Hence the detection of a QM transition induced by a GW burst from any type of  source can prove to be a probe of the spatial noncommutativity.} The results obtained in this case is in sharp contrast to those obtained in \cite{cqg} since here the polarization state of the GW signal is not relevant for the detection of spatial noncommutativity. In other words, the orientation of the GW source and the detector do not play a crucial role any more to detect the spatial NC effect. This observation is a crucial result in this paper.  
%\noindent A slightly more realistic waveform can be realized if the Gaussian function in eq.(\ref{burst_waveform_Gaussian}) is modulated by some frequency $\Omega_{0}/2 \pi$ resulting in a sine-Gaussian amplitude

\noindent Lastly,  we consider a modulated Gaussian function $g(t)$ of the form
\begin{equation}
g \left(t\right) = e^{- t^{2}/ \tau_{g}^{2}}  \,  \sin \Omega_{0}t
\label{burst_waveform_sine_Gaussian}
\end{equation}
which {\bf{represents}} a more realistic model of the GW burst signal. The Fourier transform of the said function reads
\begin{eqnarray}
\tilde{g} \left( \Omega \right) = 2 \pi \int_{-\infty}^{+\infty} g(t) e^{i \Omega t} d \Omega= \frac{i \sqrt{\pi} \tau_{g}}{2} \left[ e^{- \left(\Omega - \Omega_{0}\right)^{2}\tau_{g}^{2}/4} - e^{- \left(\Omega + \Omega_{0}\right)^{2}\tau_{g}^{2}/4} \right].
\label{sine-Gaussian_Fourier}
\end{eqnarray}
Substituting the waveform defined in eq.(\ref{sine-Gaussian_Fourier}) in eq.(s)(\ref{gwbta1}, \ref{gwbta2}), we get the transition amplitudes to be 
\begin{eqnarray}
C_{0\rightarrow 2^{(1)}}&= & \left[ e^{- \left(2 \varpi_{+} - \Omega_{0}\right)^{2}\tau_{g}^{2}/4} - e^{- \left(2 \varpi_{+} + \Omega_{0}\right)^{2}\tau_{g}^{2}/4} \right]\times  \left( 2 \sqrt{\pi}  f_{0} \varpi_{+} \tau_{g} \right) \left(A(\Lambda) \varepsilon_{+}  + B(\Lambda) \varepsilon_{\times} \right)
\label{gwptt1}
\end{eqnarray}
\begin{eqnarray}
C_{0\rightarrow 2^{(2)}}&= & \left[ e^{- \left(2 \varpi_{+} - \Omega_{0}\right)^{2}\tau_{g}^{2}/4} - e^{- \left(2 \varpi_{+} + \Omega_{0}\right)^{2}\tau_{g}^{2}/4} \right] \times  \left( 2 \sqrt{\pi}  f_{0} \varpi_{+} \tau_{g} \right) \left(C(\Lambda) \varepsilon_{+}  + D(\Lambda) \varepsilon_{\times} \right).
\label{gwptt2}
\end{eqnarray}
Therefore the corresponding transition probabilities are
\begin{eqnarray}
P_{0\rightarrow 2^{(1)}} & = & \left[ e^{- \left(2 \varpi_{+} - \Omega_{0}\right)^{2}\tau_{g}^{2}/4} - e^{- \left(2 \varpi_{+} + \Omega_{0}\right)^{2}\tau_{g}^{2}/4} \right]^{2} \times  \left( 2 \sqrt{\pi}  f_{0} \varpi_{+} \tau_{g} \right)^{2} \left(A(\Lambda)^2 \varepsilon_{+}{}^{2}  + B(\Lambda)^{2} \varepsilon_{\times}{}^{2} \right) 
\label{trans_prob_sine_Gaussian_burst}\nonumber\\
P_{0\rightarrow 2^{(2)}} & = & \left[ e^{- \left(2 \varpi_{-} - \Omega_{0}\right)^{2}\tau_{g}^{2}/4} - e^{- \left(2 \varpi_{-} + \Omega_{0}\right)^{2}\tau_{g}^{2}/4} \right]^{2} \times  \left( 2 \sqrt{\pi}  f_{0} \varpi_{-} \tau_{g} \right)^{2} \left(C(\Lambda)^2 \varepsilon_{+}{}^{2}  + D(\Lambda)^{2} \varepsilon_{\times}{}^{2} \right).
\label{trans_prob_sine_Gaussian_burst1}
\end{eqnarray}
There is a point to be noted from eq.(\ref{trans_prob_sine_Gaussian_burst}). At the low operating frequency of the detector, that is in the sub-Hz bandpass region, the two exponential terms in the transition amplitudes are almost equal and hence cancel each other. Therefore the transition amplitudes are reduced considerably. But for $2\varpi_{\pm}-\Omega_{0}= \Delta \varpi_{\pm}$, the conditions $\frac{\Delta \varpi_{\pm}}{\varpi_{\pm}}<<1$ leads to the following  transition amplitudes  
%again the physical range of the frequency $(0<\varpi> \infty)$ allows to pick up only the appropriate Fourier mode by the harmonic oscillator. If the operating frequency of the detector is low, the two exponential terms are of same size and nearly cancel each other, reducing the transition probability. The other extreme is when $2 \varpi_{+} - \Omega_{0} = \Delta \varpi_{+}$ and  $2 \varpi_{-} - \Omega_{0} = \Delta \varpi_{-}$ with $\frac{\Delta \varpi_{+}}{\varpi_{+}} << 1$ and $\frac{\Delta \varpi_{-}}{\varpi_{-}} << 1$ respectively. Then the first exponential terms are sizable while the second terms are negligible in comparison and we get the transition amplitudes 
\begin{eqnarray}
P_{0\rightarrow 2^{(1)}}  \approx   e^{- \left(\Delta \varpi_{+} \right)^{2}\tau_{g}^{2}/2}  \left(2 \sqrt{\pi}  f_{0} \varpi_{+} \tau_{g} \right)^{2} \left(A(\Lambda)^2 \varepsilon_{+}{}^{2}  + B(\Lambda)^{2} \varepsilon_{\times}{}^{2} \right)
\nonumber\\ 
P_{0\rightarrow 2^{(2)}}  \approx   e^{- \left(\Delta \varpi_{-} \right)^{2}\tau_{g}^{2}/2}  \left(2  \sqrt{\pi}  f_{0} \varpi_{-} \tau_{g} \right)^{2} \left( C(\Lambda)^2\varepsilon_{+}{}^{2}  + D(\Lambda)^{2} \varepsilon_{\times}{}^{2} \right)
\label{trans_prob_sine_Gaussian_burst_2}
\end{eqnarray} 
where the second exponential term in eq.(\ref{trans_prob_sine_Gaussian_burst}) is negligible with respect to the first one. Using the relation $\tau_{g}= \frac{2\pi}{\Omega_{max}}$, we get $\Delta\varpi_{\pm} \sim \left( \frac{2\pi \Delta\varpi_{\pm} }{\Omega_{max}}\right)^2$. Since in a bar detector $\Delta\varpi_{\pm}$ is the order of a few Hz, whereas $\Omega_{max}$ lies in the KHz range for GW burst, we see that for the more realistic sine-Gaussian function (\ref{burst_waveform_sine_Gaussian}), the transition probabilities are actually higher than the corresponding case where just the Gaussian function (\ref{burst_waveform_Gaussian}) was taken. 

Now owing to its small temporal duration the burst have a continuum spectrum of frequency over a broad range upto $f_{{\rm max}} \sim 1/ \tau_{{\rm g}}$ whereas the bar detector is sensitive only to a certain frequency window and blind beyond it. The results in eq.(s) (\ref{trans_prob_Gaussian_burst}, \ref{trans_prob_Gaussian_burst1}) and (\ref{trans_prob_sine_Gaussian_burst}, \ref{trans_prob_sine_Gaussian_burst1}) show that if there is a noncommutative structure of space, the NC shifted frequency $\varpi_{\pm}$ will be picked up for the transition as we have stated in our observation 1. Further the observations 2 and 3 also hold true for these results with aperiodic signals.
%%%%%%%%%%%%%%%%%%%%%%%%%%%%%%%%%%
\section{Size of the NC parameter and characteristic NC frequency}
%%%%%%%%%%%%%%%%%%%%%%%%%%%%%%%%%%
Before we conclude our paper let us provide a quick estimation of the dimensionless NC parameter $\Lambda$ and the characteristic NC frequency $\Lambda_{\theta}$ for the sake of completeness. It is easy to see from eq.(s) (\ref{Lambda}) and (\ref{Lambda_theta}) that both may assume different values depending on how the quantum mechanical HO is experimentally realized. In this paper we have pointed out that the phonon modes of a resonant bar detector which are excited by a GW signal behave as forced HO \cite{Magg}. So for reference mass and frequency the values appropriate for the fundamental phonon modes of a bar detector \cite{prd} are used. We also use the stringent upper-bound $|\theta| \approx 4 \times 10^{-40} {\rm m}^{2}$ \cite{carol} for spatial noncommutative parameter $\theta$. This yields 
\begin{eqnarray}   
\Lambda = \frac{m \varpi \theta}{2 \hbar} = 1.888 \left( \frac{m}{10^{3}{\rm kg}}\right) \left( \frac{\omega}{1{\rm kHz}}\right)~.     
\label{dim_less_NC1}     
\end{eqnarray}
This is a reassuring result since the estimated size of the NC parameter turns out to be of the order of unity in case of resonant bar detectors. This in turns gives the estimate for the characteristic NC frequency to be in the KHz range.
%\begin{eqnarray}   
%\Lambda_{\theta} = \frac{m \varpi^{2} \theta}{2 \hbar} = 1888 \left( \frac{m}{10^{3}{\rm kg}}\right) \left( \frac{\omega}{1{\rm kHz}}\right)^{2}
%\label{dim_less_NC1}     
%\end{eqnarray}
Since this is the amount by which the resonance point of a bar detector will be shifted due to spatial noncommutativity so we are hopeful that in case of a successful detection of GW in a resonant bar detector this NC shift will be noticeable.

% This is a really interesting feature in our analysis in this paper for all the different types of GW waveforms and can be indeed useful for an indirect evidence of the existence of the spatial NC structure of spacetime. 

\section{Conclusion}
We now summarize our findings. 
In this paper we have pointed out that it may be possible to use GW data from resonant bar detectors as a probe of noncommutative structure of space, which may exist in nature. As a demonstration of the same we have constructed a theoretical framework of a noncommutative quantum harmonic oscillator interacting with gravitational waves and worked out the transition rates between the ground state and first set of relevant excited states. We figure out that the transition probabilities are affected due to the presence of spatial noncommutativity. From the results the following important features have emerged.
\begin{enumerate}
\item There is a characteristic shift in the point of resonance, that is the frequency at which the resonant bar responds to the incoming gravitational wave maximally. This shift will be observationally noticeable in bar detectors and it is caused by a time-independent perturbation originating from the spatial noncommutativity which happens to remove the degeneracy of the second excited states of the HO.
%and the number of quantum mechanical transitions between the second excited states and the ground state of the two dimensional harmonic oscillator arising due to the incorporation of the time independent perturbation term (owing its origin to noncommutativity) in the Hamiltonian thereby removing the degeneracy of the second excited state of the quantum mechanical two dimensional harmonic oscillator. 
\item
The transitions from the ground state to the two degeneracy-lifted second excited states occur with unequal intensities and this difference in the intensity will also be observationally noticeable in case of a bar detector of GW. This will serve as another characteristic feature bearing evidence for the noncommutative structure of space. 
\item
We also find that for both linearly and circularly polarized GW these two NC characteristic features mentioned above prevail. So is the case for periodic as well as aperiodic signals of GW.  
\item
There are terms both linear and quadratic in the NC parameter $\Lambda$ in the transition rates. Thus in a different realization of quantum mechanical harmonic oscillator where $\Lambda$ may have smaller value compared to its value in case of resonant bar detector, the NC features may still stand out.
\item
We also observe that both the $+$ and the $\times$ polarizations includes the effect of spatial noncommutativity in the case of both periodic and aperiodic linearly polarized GW and which is in sharp contrast to the result obtained in \cite{cqg}. Therefore our investigation shows that the polarization state of the GW signal is not relevant for the detection of spatial noncommutativity. In other words, the orientation of the GW source and the detector do not play a crucial role any more to detect the spatial NC effect. This observation is a crucial result in this paper.
\end{enumerate}
%Note that since $\tau_{g} \sim2 \pi/\Omega_{\rm max} $, the quantity $\left( \Delta \varpi \tau_{g} \right)^{2} \sim \left(2 \pi \Delta \varpi /\Omega_{\rm max}\right)^{2}$. For bar detectors $\Delta \varpi$ is at least within the sensitive bandpass \cite{bar_1}, so it cannot be more than a few Hz, whereas $\Omega_{\rm max}$ is in the kHz range for GW burst of a few millisecond duration. This ensures that the Gaussian factor in eq.(\ref{trans_prob_sine_Gaussian_burst_1}) suppresses the transition probability only marginally. Interestingly, the Gaussian factor in eq.(\ref{trans_prob_Gaussian_burst}) is much stronger compared to that in eq.(\ref{trans_prob_sine_Gaussian_burst_1}), since $\varpi \tau_{g} >>  \Delta \varpi \tau_{g}$. Thus for the more realistic sine-Gaussian template, the transition probability is actually higher. 
Though to date none of the operational resonant bar detectors in IGEC (International Gravitational Event Collaboration) \cite{bar_detectors_1, bar_detectors_2, bar_detectors_3, bar_detectors_4, bar_detectors_5} have yet achieved the required quantum limit of sensitivity for a successful detection of GW, but a lot of work is currently being done with the new generation of spherical resonant detectors MiniGrail \cite{MiniGrail} and Schenberg \cite{Schenberg}. We would like to conclude with the statement that the considerations in the present paper suggest that the operation of these resonant detector groups may possess the potential to establish the possible existence of a granular structure of our space as a by-product in the event of a direct detection of gravitational wave. Search in this direction should be carried out as this would then be a fundamental discovery about the mysteries of nature.

%-----------al positio--------------------------
\section*{Acknowledgment} \noindent AS and SG would like to thank Inter University Centre for Astronomy $\&$ Astrophysics, Pune for the Visiting Associateship programme. SB would like to thank the Government of West Bengal for financial support. AS acknowledges the financial support of DST SERB under Grant No. SR/FTP/PS-208/2012. SG acknowledges the support by DST-SERB under the Start Up Research Grant (Young Scientist), File No.YSS/2014/000180. 
%-------------------------------------
%\noindent  where a considerable part of this work was
%completed. The authors would also like to thank the referee for
%useful comments.

%%%%%%%%%%%%%

%%%%%%%%%%%%%%%%%%%%%%%%%%%%%%%%%%%%%%%%%%%%%%%%%%%%%%%%%%%%%%%%%%%%%%%%%%%%%%%%
\end{document}